\documentclass[pre, preprint, showpacs, endfloats]{revtex4-1}
\usepackage{natbib}
\usepackage{amsmath}
\usepackage{graphicx}
\usepackage{verbatim}
\usepackage{subfigure}
\usepackage{hyperref}
\usepackage{xcolor}

\newcommand{\tb}[1]{\ensuremath{\mathbf{#1}}}
\newcommand{\markthis}[1]{#1}

\begin{document}

\title{The nucleation and propagation of solitary Schallamach waves}
\author{Koushik Viswanathan}
\email[E-mail: ]{kviswana@purdue.edu}
\author{Anirban Mahato}
\author{Srinivasan Chandrasekar}
\affiliation{Center for Materials Processing and Tribology\\Purdue University, West Lafayette, IN 47907-2023}
\date{\today}

\begin{abstract}
We isolate single Schallamach waves --- detachment fronts that mediate inhomogeneous sliding between an elastomer and a hard surface --- to study their creation and dynamics. Based on measurements of surface displacement using high--speed \emph{in situ} imaging, we establish a Burgers vector for the waves. The crystal dislocation analogues of nucleation stress, defect pinning and configurational force are demonstrated. It is shown that many experimentally observed features can be quantitatively described using a conventional model of a dislocation line in an elastic medium. We also highlight the evolution of nucleation features such as surface wrinkles, with consequences for interface delamination.
\end{abstract}


\maketitle


\section{Introduction}
\label{sec:introduction}

While the phenomenological study of friction between solid surfaces has a long history \cite{BowdenTabor_Friction}, it was not until the mid-twentieth century that the microscopic aspects were first probed \cite{Merchant_JApplPhys_1940, *BowdenTabor_Nature_1942}. Subsequently, significant attention has been devoted to the fundamental microscale mechanisms underlying the phenomena of static and dynamic friction \cite{GerdeMarder_Nature_2001, RubinsteinETAL_Nature_2004, Vallette_Gollub_PhysRevE_1993}. Simple universal laws such as those of Amontons and Coulomb, though used extensively, frequently remain unsatisfactory \cite{BaumbergerCaroli_AdvPhys_2006}. For example, under extreme sliding conditions, the friction depends on the area of contact \cite{Orowan_ProcInstMechEngg_1943, *vonKarman_ZAMM_1925}. Likewise, at low sliding velocities, sliding friction is non--trivially dependent on velocity \cite{Persson_PhysRevB_2001} and normal load history \cite{RanjithRice_JMechPhysSolids_2001, *Caroli_PhysRevE_2000}, resulting in the occurence of inhomogeneous modes of sliding with localized slip.

A soft adhesive elastomer sliding on a smooth surface is a model system that exhibits inhomogeneous sliding modes, while also capturing the physics underlying processes of practical and industrial interest \cite{Johnson_ContactMechanics_1987}. At a length scale of a few hundred micrometers and low relative sliding velocity ($\lesssim 10$ mm/s), motion between the two surfaces does not occur homogeneously, but via the propagation of \lq waves of detachment\rq , also known as Schallamach waves \cite{Schallamach_Wear_1971, Barquins_Wear_1985, RandCrosby_ApplPhysLett_2006}. Under similar conditions, sometimes another inhomogeneous mode of sliding, called the self--healing slip pulse is also observed \cite{*[{See }] [{ and references therein.}] BaumbergerETAL_PhysRevLett_2002}. Schallamach waves have been likened to crystal dislocations \cite{Gittus_PhilMag_1975, *BriggsBriscoe_PhilMag_1978} or rucks in carpets \cite{VellaETAL_PhysRevLett_2009, *KolinskiETAL_PhysRevLett_2009} \cite{*[{The analogy between carpet rucks and dislocations causing slip in crystals has often been attributed in the recent literature to E. Orowan. However, it appears to have been proposed by Bragg --- see the article by W. Lomer in }][{, pp. 115 -- 118.}] PhillipsThomas_BraggLegacy_1990}. Such comparisons have only been qualitative; however, they have helped rationalize some observed features such as the locality of surface slip and existence of a nucleation stress.

Motivated by these considerations, we have further explored, using experiments, the similarity between a single Schallamach wave and a dislocation line in an elastic medium. For this purpose, a long adhesive contact was established between an elastomer and a solid surface, enabling observation of solitary Schallamach waves (wave pulses). This provides a suitable framework in which to study their characteristics quantitatively. High--speed \emph{in situ} imaging was used to capture their nucleation and propagation, at resolution of $\sim$ 2 $\mu$m.

Improved observations of the intrinstic features of isolated Schallamach waves should help better understand inhomogeneous interfacial sliding phenomena prevalent in earthquake ruptures \cite{Heaton_PhysEarthPlanetInt_1990}, polymer friction \cite{RubinsteinETAL_Nature_2004, Vallette_Gollub_PhysRevE_1993} and locomotion of soft--bodied animals \cite{Trueman_SoftBodiedLocomotion_1975}. It could also shed new light on the relation between Schallamach waves and the self--healing slip pulse mode of sliding \cite{BaumbergerETAL_PhysRevLett_2002}.

The paper is organized as follows. Details of the experimental setup are provided in Sec.~\ref{sec:experimental}, followed by high--speed photographic observations of Schallamach wave nucleation and propagation (Sec.~\ref{sec:results}). Using image analysis techniques, we obtain quantitative information about the individual wave properties that help explain various features of the nucleation and propagation stages. The results are analyzed in detail in Sec.\ref{sec:analysis}, along with a discussion of why inhomogeneous sliding modes occur. The principal findings are summarized in Sec.~\ref{sec:conclusions}.


\section{Experimental details}
\label{sec:experimental}

\begin{figure}[ht]
\centering
\includegraphics[scale=0.65]{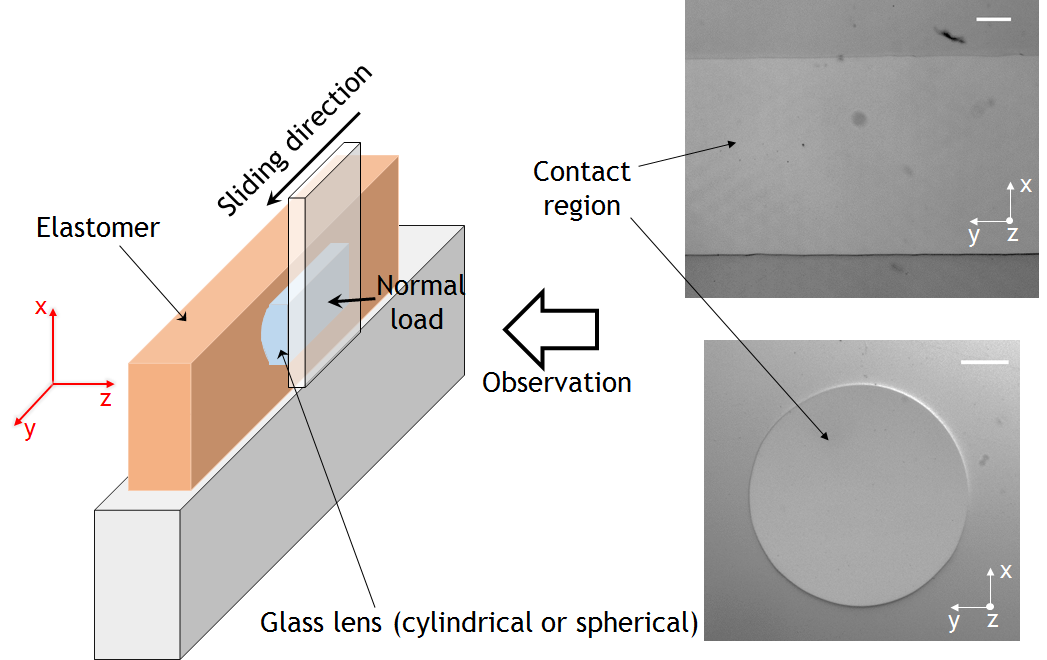}
\caption{(Color online) Schematic of the experimental setup, with reference coordinates. Images on the right show the contact regions for the cylindrical (top) and spherical (bottom) lens geometries. Scale bar corresponds to 200 $\mu$m.}
\label{fig:sphereCylinderSchematic}
\end{figure}

The adhesive contact used is that between an uncoated plano--convex lens, made of synthetic glass (Edmund Optics) and the elastomer PolyDiMethylSiloxane (PDMS, Sylgard 184 from Dow Corning). A schematic of the experimental setup is shown in Fig.~\ref{fig:sphereCylinderSchematic}. The elastomer and the lens are optically transparent. The contact region is illuminated by a 120 W halogen lamp and observed by a microscope (Nikon Optiphot) mounted in front of a high--speed camera (PCO dimax). This system was used to image the contact region at framerates of $5000 - 8000$ Hz. The resulting spatial resolution was $1.9 - 2.8$ $\mu$m per pixel, depending on the microscope lens used. Normal and tangential forces in sliding were measured using a piezoelectric dynamometer (Kistler). 

Two different plano--convex lens geometries were used for the experiments --- a spherical lens of radius $R_s$ = 5 mm and a cylindrical lens of radius $R$ = 16.25 mm and length $L$ = 25 mm. Sample images of the contact region with the cylindrical lens (Fig.~\ref{fig:sphereCylinderSchematic} (right, top)) and the spherical lens (Fig.~\ref{fig:sphereCylinderSchematic} (right, bottom)). In both cases, the size of the contact region was maintained constant, between experiments. For the cylindrical lens geometry, the angle subtended by the contact region at the lens axis was around $1.7^\circ$; the lens curvature caused a shift in the image of $\simeq 1$ $\mu$m, which was, however, not resolvable by the imaging system. Upon contact, the cylindrical lens formed a long aspect--ratio adhesive \lq channel\rq\ in which to propagate solitary Schallamach waves. From the high speed observations, this was found to be most conducive for the production of single wave pulses at the interface. Characteristics of wave nucleation were studied using the spherical lens geometry, because the resulting finite contact region allowed complete observation of the contact edges.

The PDMS elastomer sample was prepared by mixing a base (vinyl-terminated polydimethylsiloxane) with a curing agent (methylhydrosiloxane--dimethylsiloxane copolymer) in the ratio 10:1 by weight. The resulting mixture was cured for 12 hours at $60^\circ$ C. The PDMS was cast in a mold into slab type specimens, with dimensions of $L = 70$ mm $\times$ $H = 22$ mm in the $xy$ plane (see Fig.~\ref{fig:sphereCylinderSchematic}). $L$ is the length in the sliding direction. The thickness of the slab was 12 mm. For experiments requiring a longer sliding length, a slab with $L = 90$ mm was used. The Young's modulus and Poisson's ratio for PDMS were estimated from reported shear \cite{Mark_PolymerDataHandbook_2009} and bulk modulus \cite{SchmidMichel_Macromolecules_2000} values for Sylgard 184 as $E \simeq 800$ kPa and $\nu = 0.45$ respectively. The sample was mounted on a linear slide which could impose sliding velocities of 10 $\mu$m/s -- 20 mm/s. The camera and the indenter (lens) were fixed on rigid supports as shown in Fig.~\ref{fig:sphereCylinderSchematic} (left). 

The starting contact size was kept constant in all the experiments by maintaining the applied initial normal load at 35 mN (spherical lens) and 55 mN (cylindrical lens). The dynamics of the interface maybe expected to vary with the normal load. 

The high--speed image sequences obtained in the experiments were analyzed to obtain displacements for each pixel between successive images. This is done by assuming that the image intensity is convected with the physical velocity field. Since the framerate is kept constant, the inter--frame displacement is proportional to the instantaneous velocity. A brief description of the image processing methods is provided in Appendix \ref{sec:appendix}. Once the velocities of every pixel are obtained for each image, specific surface properties are determined by following a set of predetermined \lq virtual\rq\ tracer points. These are pixel locations in the first image (with perfect adhesive contact) spaced 2 pixels apart, along the horizontal contact mid--line ($x=0$) in the middle of the contact region. Their positions $(x_i, y_i)$ are altered between successive frames, using the local velocity field determined for that particular frame. All of the results are presented and discussed in the elastomer rest--frame. It is in this sense that the \lq leading\rq\ and \lq trailing\rq\ edges of the lens are defined. 


\section{Observations}
\label{sec:results}

Interfacial slip via Schallamach waves consists of two main stages --- nucleation of a wave and its propagation through the interface. These are studied using the spherical and cylindrical lens geometries respectively. The former is chosen due to the small circular contact zone, which enables easier observation. The latter removes the effects of contact geometry, thereby allowing us to observe the intrinsic propagation characteristics of individual Schallamach wave pulses. 

\subsection{Nucleation of Schallamach waves}

\begin{figure}
\clearpage
  \centering
  \mbox
  {
    \subfigure[\label{fig:nucleationA}]{\includegraphics[scale=0.65]{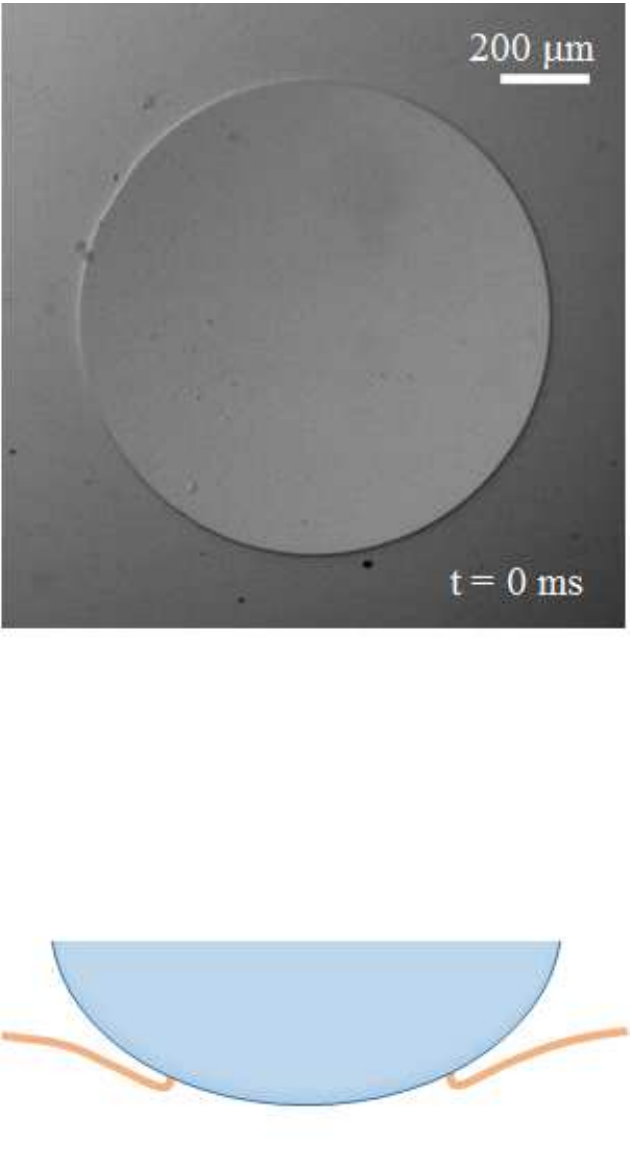}}
    \subfigure[\label{fig:nucleationB}]{\includegraphics[scale=0.65]{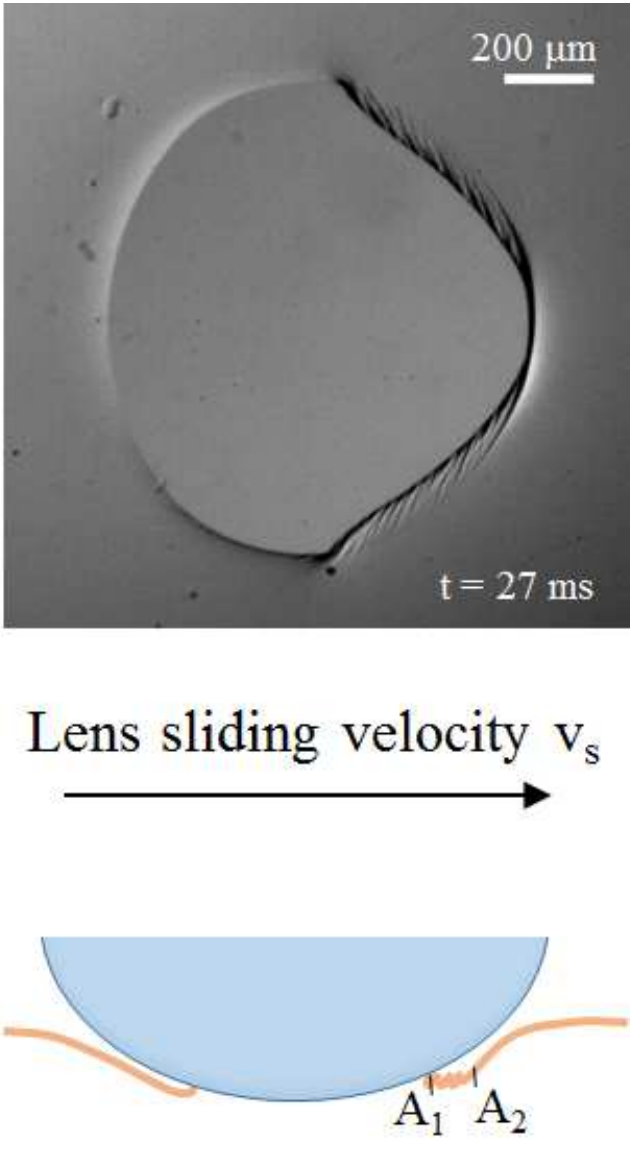}}
    \subfigure[\label{fig:nucleationC}]{\includegraphics[scale=0.65]{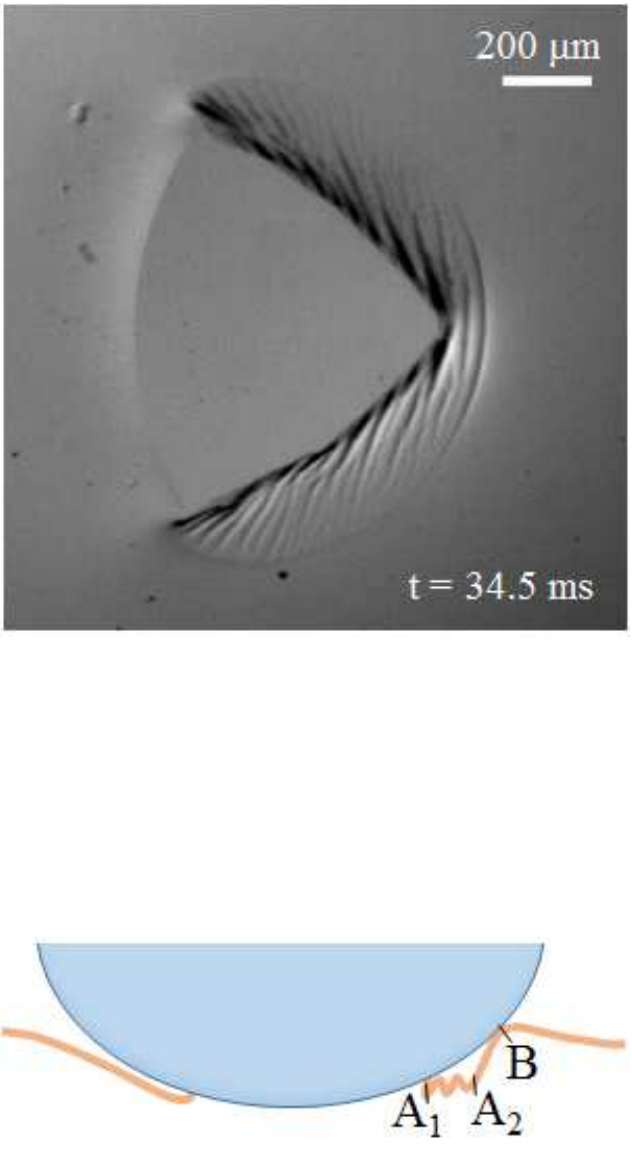}}
    \subfigure[\label{fig:nucleationD}]{\includegraphics[scale=0.65]{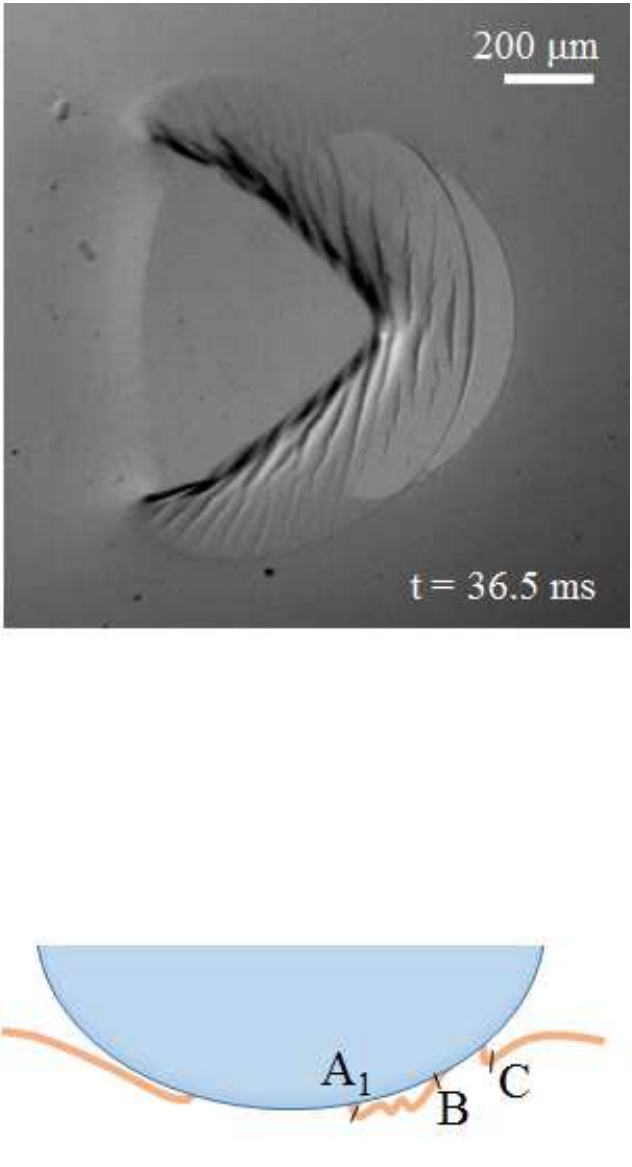}}
  }
  \caption{(Color online) Four frames from a high--speed sequence showing the nucleation of Schallamach waves (top row) with a schematic side--view (bottom row). (a) Circular contact region before application of tangential force. (b) Change in shape of contact region and accompanying buckling instability the initiates wave nucleation. (c) The surfaces readhere ahead of the lens (point $B$). (d) A single Schallamach wave pulse travels through the contact region. $v_s = 20$ mm/s, spherical lens.}
  \label{fig:nucleation_result}
\end{figure}

Schallamach waves nucleate due to a buckling instability of the elastomer surface \cite{Schallamach_Wear_1971, Barquins_Wear_1985}. A prototypical nucleation event is shown in Fig.~\ref{fig:nucleation_result} (top row) with the corresponding schematic side--view in Fig.~\ref{fig:nucleation_result} (bottom row). Initially, in Fig.~\ref{fig:nucleationA}, the spherical lens and the elastomer surface are in adhesive contact. When a relative sliding velocity is imposed, the elastomer free surface ahead of the lens is compressed, causing it to buckle (Fig.~\ref{fig:nucleationB}). This compression results from a combination of the applied tangential force and adhesion at the interface.

In order to maintain a constant sliding velocity $v_s$, continued application of the tangential force is necessary. The van der Waals force between the surfaces causes the elastomer to reattach to the lens, at point $B$ in Fig.~\ref{fig:nucleationC}. An air pocket (region $A_1 B$) is thus trapped inside the contact region. The presence of a strong shear stress gradient causes this pocket to traverse the length of the contact region in the form of a single Schallamach wave, as seen in Fig.~\ref{fig:nucleationD}. The region $BC$ in the wake of the wave is now again in adhesive contact. These waves hence travel from the leading edge of the lens to the trailing edge i.e. in a direction opposite to the imposed sliding velocity $v_s$.

\begin{figure}
  \centering
  \includegraphics[scale=0.75]{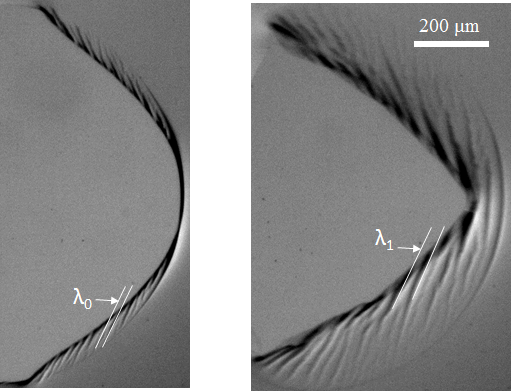}
  \caption{Wrinkles accompanying the nucleated wave. (Left) Initial pattern, with wrinkle wavelength $\lambda_0 = 18\text{ }\mu\text{m}$. (Right) Increase in wrinkle wavelength to $\lambda_1 = 40\text{ }\mu\text{m}$ due to continued application of tangential stress.}
  \label{fig:wrinkles}
\end{figure}

A prominent feature of Fig.~\ref{fig:nucleation_result} is the wrinkle pattern on the surface (region $A_1 A_2$) accompanying the wave \cite{*[{The only report of a reference to wrinkles that we are aware of is }][{}] KoudineBarquins_JAdhesionSciTech_1996}. This is seen in the movie M1 (supplemental material \cite{SuppMat}) and shown in Fig.~\ref{fig:wrinkles}. These wrinkles are also compression--induced features, akin to the formation of sulci \cite{HohlfeldMahadevan_PhysRevLett_2011}, and have important consequences for wave propagation. The average spacing between adjacent wrinkles gives the pattern wavelength. The initial value of the wavelength $\lambda_0$, as measured from the images, is $18$ $\mu$m, see Fig.~\ref{fig:wrinkles} (left). Upon further application of shear, even though the two surfaces remain adhered at the interface (Figs.~\ref{fig:nucleationB} and \ref{fig:nucleationC}), there is an increase in the compressive stress on the elastomer free surface. The pattern wavelength correspondingly increases to $\lambda_1 = 40\text{ }\mu \text{m}$ (Fig.~\ref{fig:wrinkles} (right)), with is approximately twice the initial value $\lambda_0$. There is an accompanying increase in wrinkle amplitude. 

For low sliding velocity ($v_s \sim 5 \text{ mm/s}$ for the spherical lens), where the interfacial shear stress is not sufficient to change the wavelength, large amplitude wrinkles do not form on the surface, see Movie M1 \cite{SuppMat}. In this regard, it is interesting that an elastic film on an elastomer substrate, under longitudinal compression --- a similar loading condition as in the present experiments --- also exhibits surface wrinkling as well as a period doubling instability for large loads \cite{BrauETAL_NaturePhys_2010}. The image corresponding to Fig.~\ref{fig:nucleationC} represents the end of the nucleation of a single Schallamach wave; the wrinkles subsequently move in consonance with it (Fig.~\ref{fig:nucleationD}).

Another important feature of the wave nucleation is the shape of the contact region in the sequence in Fig.~\ref{fig:nucleation_result}. The leading edge of the initially circular contact region is stretched in the forward direction due to the applied force. As this force is increased, the inclination of the leading contact edge increases until nucleation is complete (Fig.~\ref{fig:nucleationC}), following which the nucleated wave maintains its profile (Fig.~\ref{fig:nucleationD}). 

\subsection{Propagation through the interface}

Once a solitary Schallamach wave is nucleated, it traverses the contact region, due to a stress gradient. The propagation characteristics of the nucleated wave pulse are best observed in the cylindrical contact. The initial contact region resembles a long, thin adhesive \lq channel\rq . The length of the contact region $L = 2.5$ cm and width $2a \sim 1$ mm were kept constant for all the experiments. The images were recorded in the middle of the contact but were found to be consistent along the entire length.

\begin{figure}
\clearpage
\centering
\includegraphics[scale=0.68]{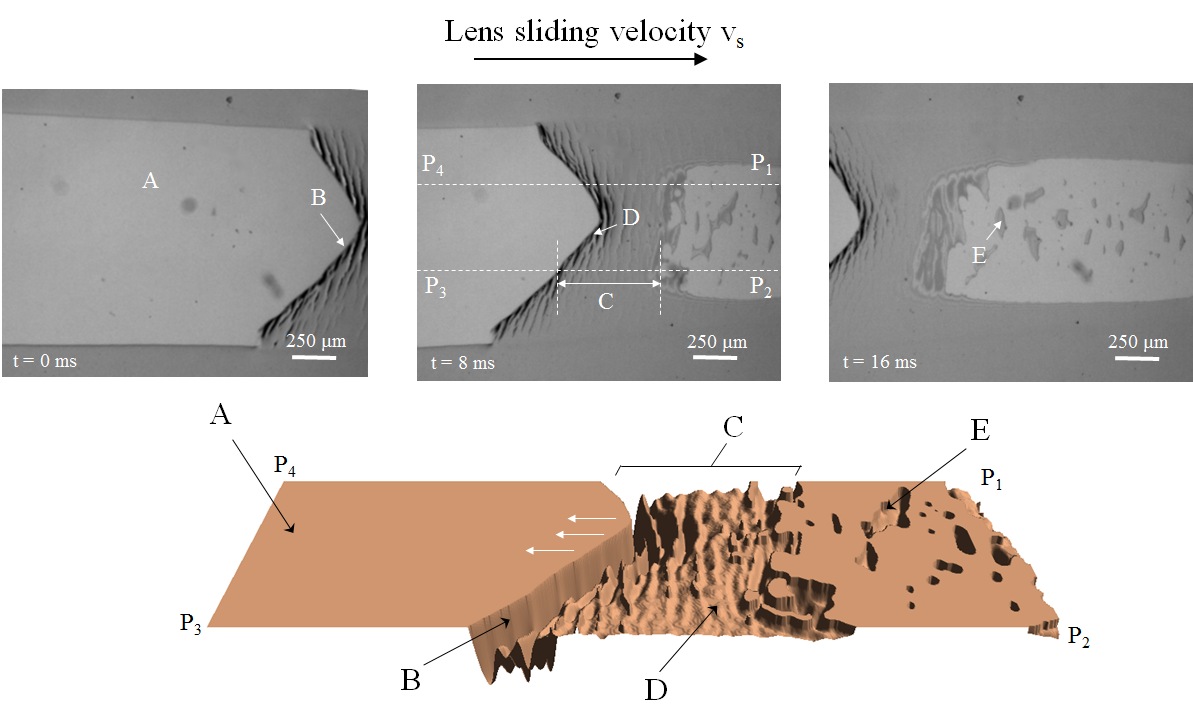}
	\caption{(Color online) Schallamach wavefront and its propagation features. (Top row) Sequence of images showing propagation of a solitary Schallamach wave. The full length of the contact region is about 20X the length shown in the images. (Bottom) 3D intensity plot, derived from the images, showing various features on the elastomer surface. $A$ - Initial adhesive contact between the surfaces, $B$ - front of Schallamach wave, $C$ - Extent of trapped air pocket comprising the wave, $D$ - Wrinkles on the surface and $E$ - Incomplete readhesion after wave passage. \markthis{Wave velocity $v_w \simeq 110$ mm/s}. $v_s = 2.5$ mm/s, cylinder lens.}
	\label{fig:timeFrames}
\end{figure}

A sequence of frames from the cylinder lens contact is shown in Fig.~\ref{fig:timeFrames} (top row). The image intensity is depicted in 3D in Fig.~\ref{fig:timeFrames} (bottom row) and follows the elastomer surface profile. The elastomer and lens are initially in perfect contact (region $A$) with the Schallamach wavefront clearly demarcated (edge $B$). The wave itself is seen as a depression (region $C$) due to the trapped air pocket. The surface wrinkles (eg. at point $D$) are also visible. Once the solitary Schallamach wave has passed, readhesion between the surfaces is incomplete, leaving small stationary residual air pockets (like at point $E$). In movie M2 \cite{SuppMat}, it is seen that these air pockets form exactly over the free surface wrinkle, pattern owing to increased strain concentration in the wrinkles. Such wrinkles, formed during wave nucleation, were found to cause significant interface delamination after the passage of successive Schallamach wave pulses. 

\begin{figure}
  \centering
  \mbox
  {
    \subfigure[\label{fig:propagationA}]{\includegraphics[scale=0.65]{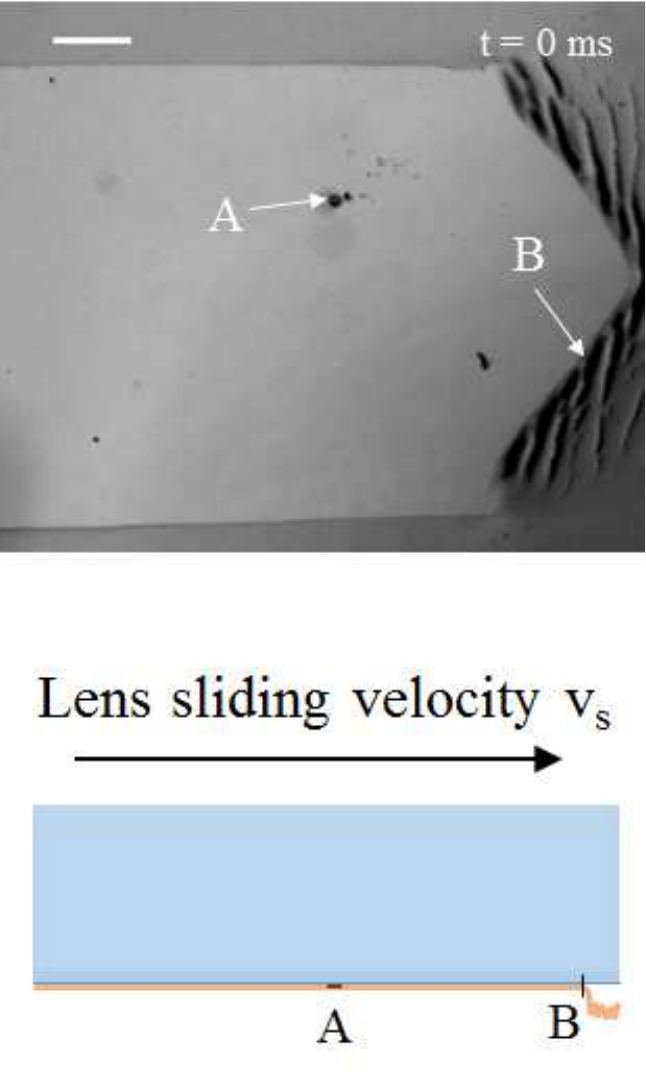}}
    \subfigure[\label{fig:propagationB}]{\includegraphics[scale=0.65]{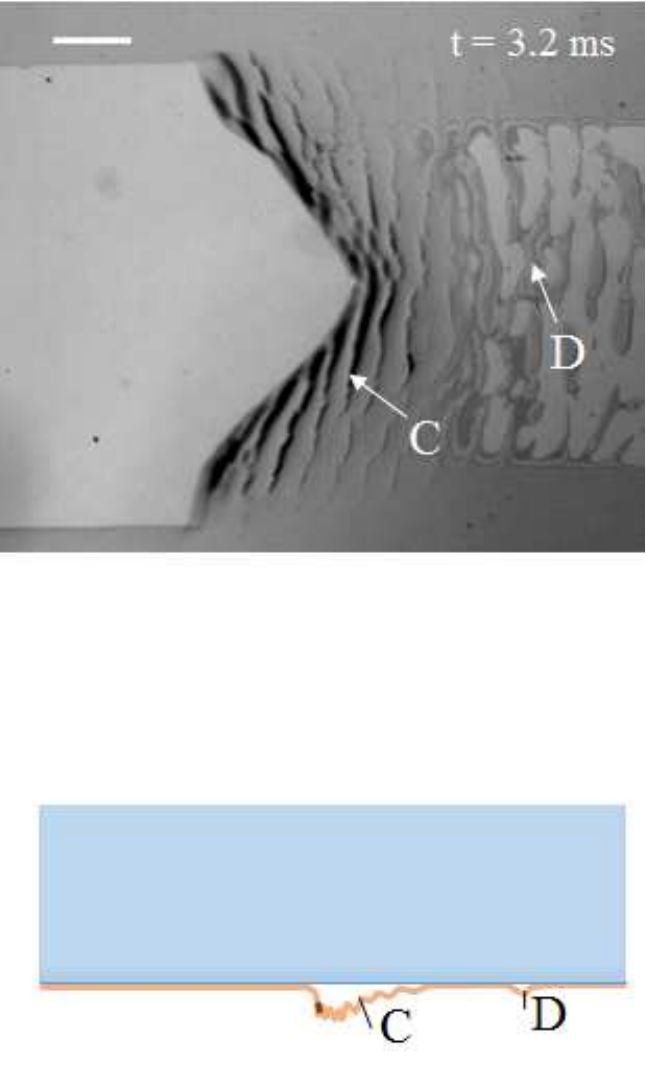}}
    \subfigure[\label{fig:propagationC}]{\includegraphics[scale=0.65]{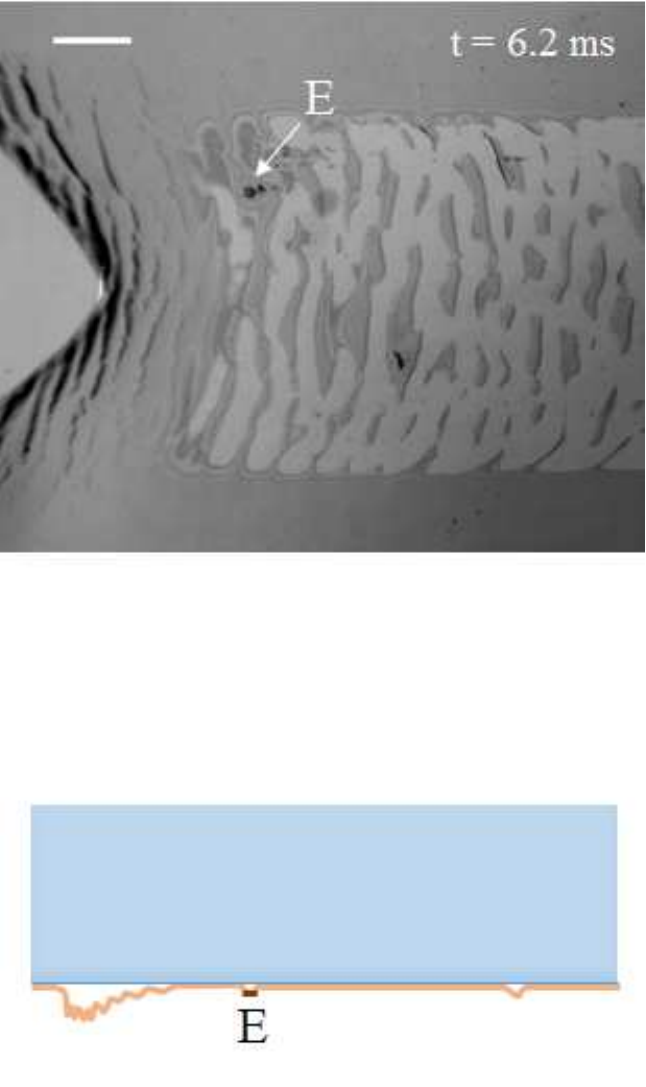}}
    \subfigure[\label{fig:propagationD}]{\includegraphics[scale=0.65]{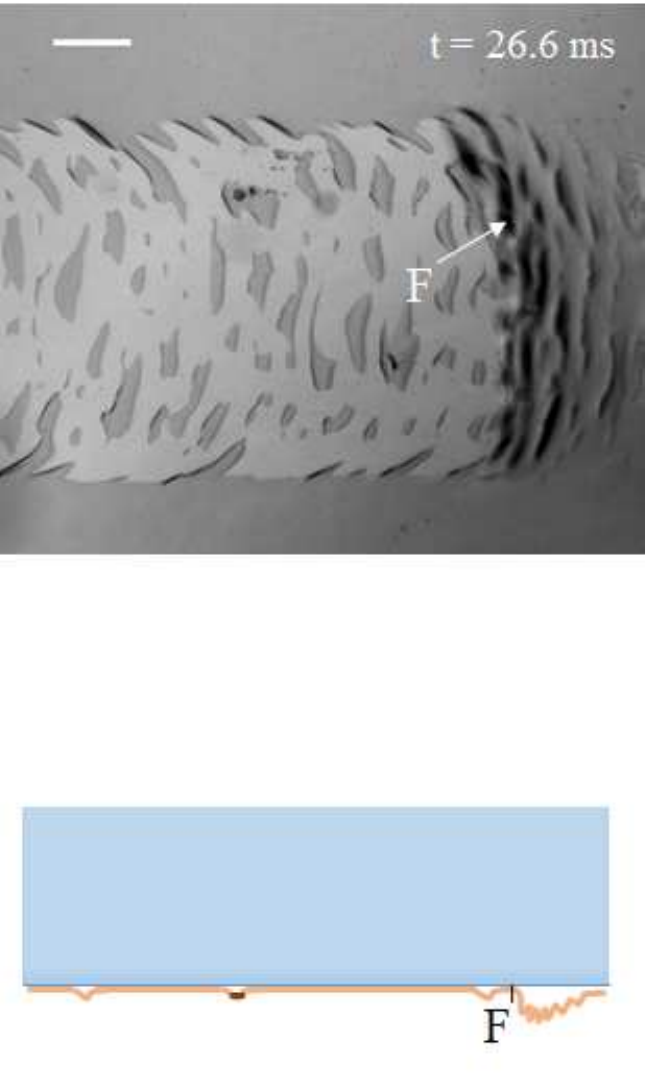}}
  }
  \caption{(Color online) Propagation of single Schallamach wave pulses in an adhesive contact. The waves retain their profile over long distances. The length of the contact region is 2.5 cm. Scale bar corresponds to $250$ $\mu$m; Wave velocity $v_w \simeq 380$ mm/s. $v_s = 10$ mm/s, cylinder lens}
  \label{fig:propagation_result}
\end{figure}

The result of passage of a single Schallamach wave is also brought out by its effect on the surface displacement of a dirt particle. Fig.~\ref{fig:propagationA} shows perfect adhesive contact between the cylinder lens and elastomer surface, with a dirt particle (point $A$) on the elastomer surface. As in Fig.~\ref{fig:timeFrames}, the leading front of the incoming wave pulse is denoted by $B$. When the wave propagates through the observed region (Fig.~\ref{fig:propagationB}), the wrinkles $C$ again result in incomplete readhesion in the wake (point $D$). The dirt particle $A$ is displaced from its original position as the wave pulse passes over it. A small air pocket is also left around it in the process (point $E$ in Fig.~\ref{fig:propagationC}). From the initial and final positions of the particle at $A$ in Fig.~\ref{fig:propagationA}, it is clear that relative motion between the surfaces has occured intermittently and only due to wave passage, see movie M2 \cite{SuppMat}. This cycle is repeated when a second wave ($F$ in Fig.~\ref{fig:propagationD}) is nucleated and traverses the contact region. Since the $v_s$ in Fig.~\ref{fig:propagation_result} is larger than in Fig.~\ref{fig:timeFrames}, more air pockets remain trapped in the wake of the wave. The contact condition is hence changed for subsequent wave nucleation and the shape of the following wave pulses is altered, as in Fig.~\ref{fig:propagationD}.

\begin{figure}
  \centering
  \includegraphics[scale=0.7]{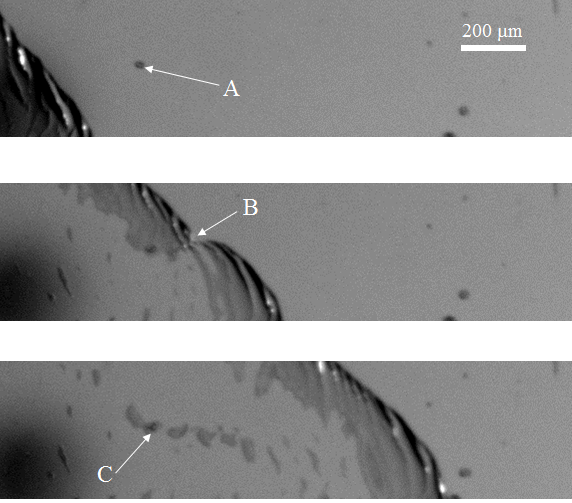}
  \caption{Pinning of a single Schallamach wave by static dirt particles. The wavefront approaches a single dirt particle $A$ (top row), and is bent by it (point $B$, middle row). The wave then regains its original shape, while leaving behind an air pocket $C$ around the dirt particle.}
  \label{fig:pinning}
\end{figure}

An interesting phenomenon is observed during propagation past stationary dirt particles attached to the lens, as shown in Fig.~\ref{fig:pinning}. Here, the wavefront approaches a static dirt particle $A$ in the contact region and gets \lq pinned\rq , causing a bend in the wave profile (point $B$). As the front moves away from the particle, the wave regains its original profile, leaving a residual air pocket $C$ around the particle. This resembles the motion of dislocations in crystals past static obstacles (solute particles), leaving behind so-called Orowan loops \cite{Nabarro_CrystalDislocations_1967}. For dislocations in metals, this is known to lead to the Fisher, Hart and Pry (FHP) effect \cite{Cottrell_MechPropMatter}. Under dilute solute particle concentration, the dislocation line is bent by the obstacles and it eventually relaxes its shape.

\subsection{Properties of a solitary Schallamach wave}

In studying the propagation of a single Schallamach wave, three different velocities must be distinguished --- the imposed (remote) sliding velocity $v_s$, the local material velocity $\mathbf{v}_p$ and the velocity $v_w$ of propagation of the Schallamach wave. In each sliding experiment, $\mathbf{v}_p$ and $v_w$ could, in principle, vary along the contact.

\begin{figure}
  \centering
    \subfigure[\label{fig:surfaceDisp}]{\includegraphics[scale=0.6]{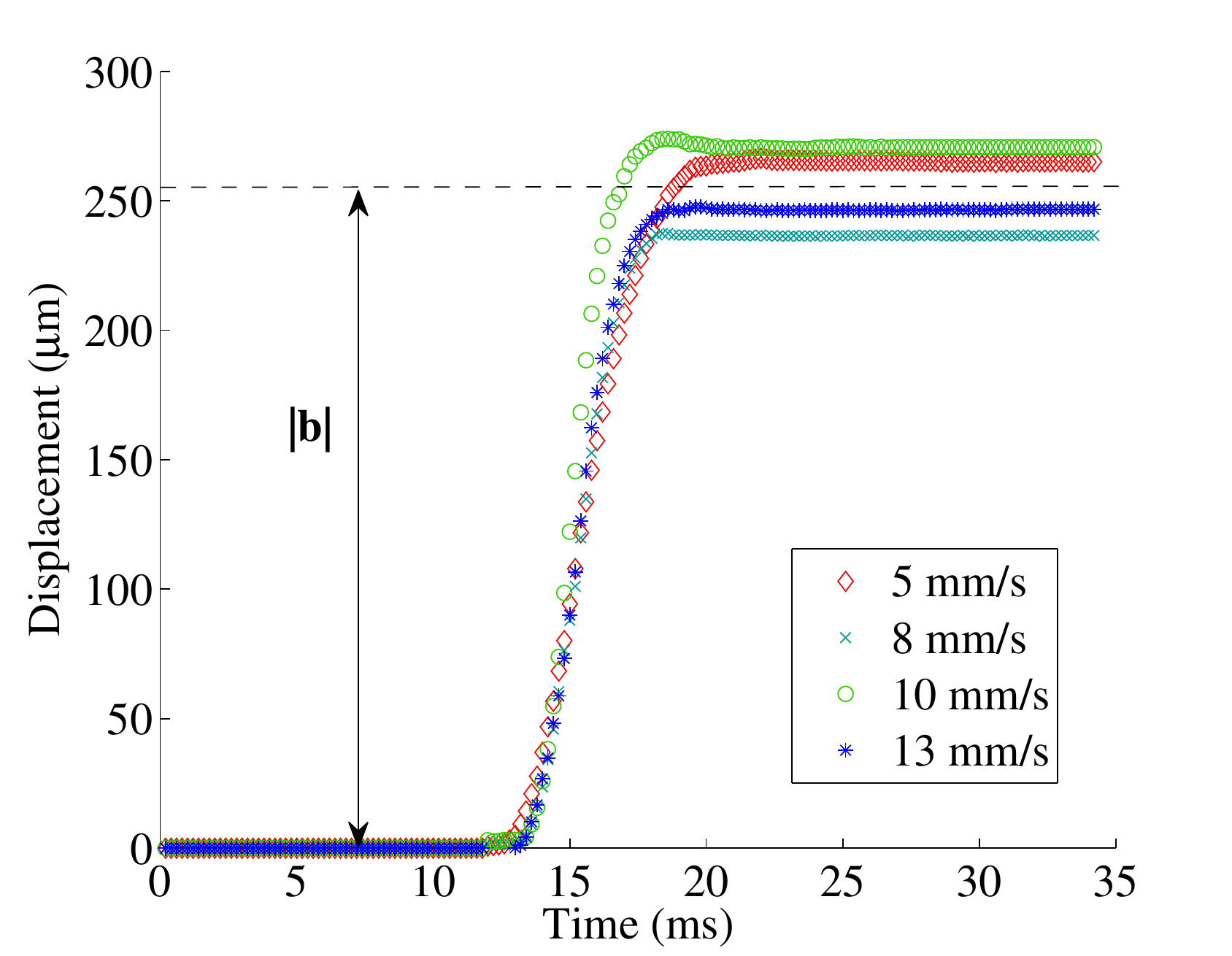}}\\
    \subfigure[\label{fig:spaceTime}]{\includegraphics[scale=0.6]{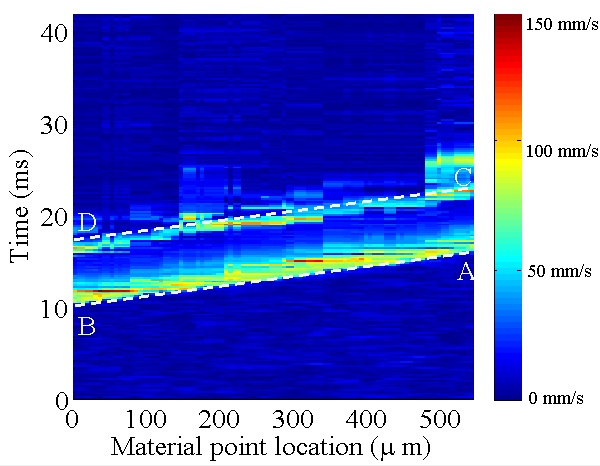}}

  \caption{(Color online) Burgers vector and group velocity of a Schallamach wave pulse. (a) Mean surface displacement due to a single Schallamach wave, for various values of $v_s$; the magnitude of the jump denotes the $|\mathbf{b}|$ of the wave. (b) Space--time diagram showing local velocity $|\mathbf{v}_p|$ for points on surface, $v_s = 2.5$ mm/s. $AB$ and $CD$ denote the front and rear of the wave pulse. Cylinder lens.}
  \label{fig:surfDisp_spaceTime}
\end{figure}

Standard image analysis techniques were applied to the high--speed image sequences of an isolated wave to obtain  the material velocity $\mathbf{v}_p(x,y)$ for each pixel $(x,y)$ in an image frame (See Appendix \ref{sec:appendix} for details). By tracking a set of horizontal \lq virtual\rq\ tracer points $(x_i, y_i)$ lying in the initally perfect contact region, the relative inter--frame displacement of the surfaces during wave propagation was obtained. This is shown for four different values of $v_s$ in Fig.~\ref{fig:surfaceDisp}. The graph shows a distinct jump, implying that relative motion occurs only due to wave passage; the surfaces are otherwise stationary and in perfect contact. An analogous situation prevails during the irreversible displacement (slip) caused by motion of an edge dislocation on a crystal glide plane. In this case, the displacement magnitude is given by the dislocation Burgers vector. Similarly, the displacement jump in Fig.~\ref{fig:surfaceDisp} can be associated with a Burgers vector $\mathbf{b}$ for the solitary Schallamach wave. It is clear from Fig.~\ref{fig:surfaceDisp} that $|\mathbf{b}| =255$ $\mu$m and is independent of $v_s$. It is determined, for a given contact geometry, by the substrate material properties. Both of these characteristics are also true for the dislocation Burgers vector \cite{Nabarro_CrystalDislocations_1967}. The value of $|\mathbf{b}|$ for a Schallamach wave depends on the contact dimensions, and hence, on the normal load.

The velocity magnitudes $|\mathbf{v}_p(x_i, y_i)|$ of each of the material tracer points $(x_i, y_i)$ may now be assembled in the form of a space--time diagram, as shown in Fig.~\ref{fig:spaceTime}. The $x$-axis is the initial location of the material tracer points and $y$-axis denotes time. The pixel color values denote $|\mathbf{v}_p(x_i, y_i)|$ of each tracer point for a particular time slice. The leading edge of the Schallamach wave pulse is represented by the line $AB$ and the trailing edge by $CD$. The slopes of these lines are equal, giving a wave group velocity $v_w = 110$ mm/s and $v_w/ v_s \simeq 45$. The equal values of the slopes show that a Schallamach wave pulse maintains its shape during propagation over the long contact. In general, when $v_s$ is increased, $v_w$ also increases but $v_w/v_s$ appears to reduce a little. For the range of $v_s$ used, this ratio was between 35 and 50. Note that along the wave pulse profile, $v_w$ is different from the local phase velocity $|\mathbf{v}_p(x_i,y_i)|$ (=$50 - 140$ mm/s). It is interesting that both $v_w$ and $\mathbf{v}_p$ are much smaller than the shear wave velocity for PDMS ($\sim 15$ m/s).

\subsection{Motion of multiple Schallamach waves}
\begin{figure}
\clearpage
  \centering
  \mbox{
    \subfigure[\label{fig:forceTrace}]{\includegraphics[scale=0.55]{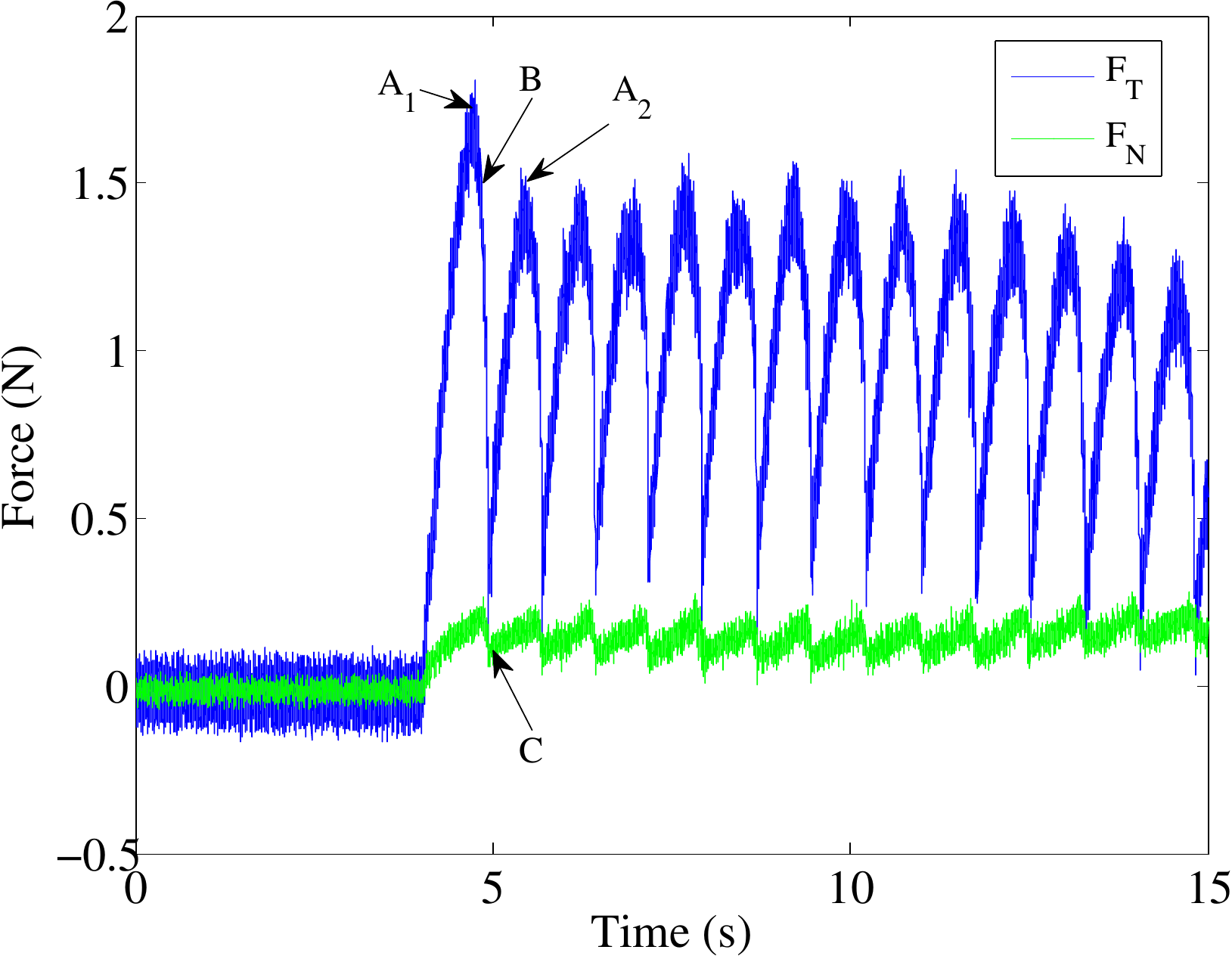}}
    \subfigure[\label{fig:genFreq}]{\includegraphics[scale=0.55]{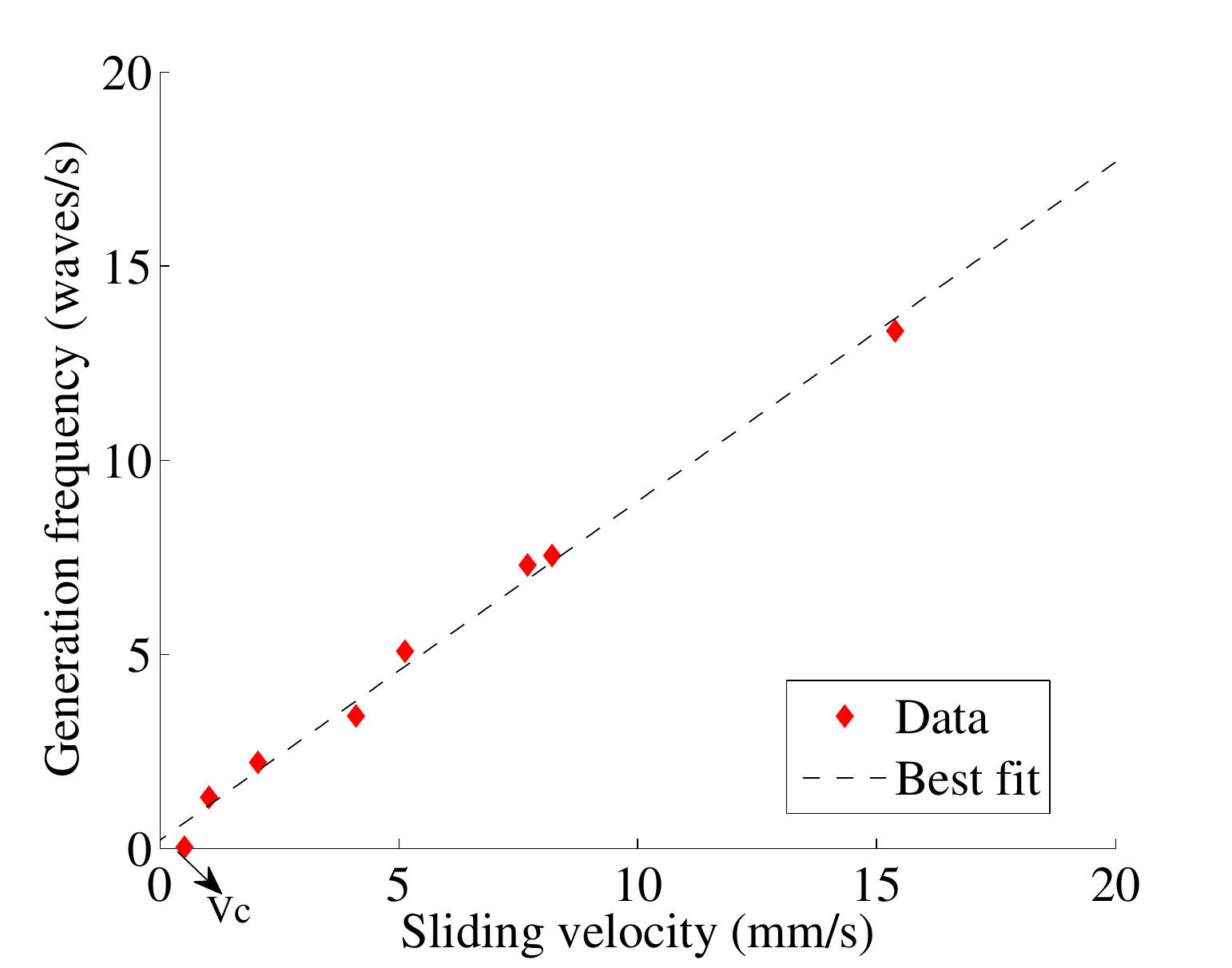}}
  }
  \caption{(Color online) Normal $(F_N)$ and tangential $(F_T)$ forces, and generation frequency ($n$) for Schallamach waves. (a) Time--variation of $F_T$ (blue or dark grey) and $F_N$ (green or light gray) for $v_s = 1$ mm/s. Each oscillation of $F_T$ represents the propagation of a single wave. Critical force for nucleation $F_C$ is marked by point $A$. (b) Dependence of $n$ on $v_s$. The critical velocity for Schallamach wave formation is $v_c = 150$ $\mu$m/s. Cylinder lens.}
  \label{fig:forceTrace_genFreq}
\end{figure}

The tangential force $F_T$ on the elastomer was measured during sliding and is shown (in blue (dark grey)) in Fig.~\ref{fig:forceTrace}. This force provides a measure of the shear stress at the interface. Since the experiments were performed under velocity control, the force varies with time. The measured value of the normal force, $F_N$ (in green (light grey)), is also seen to oscillate with time in Fig.~\ref{fig:forceTrace}. The force data were correlated with the high--speed image data to confirm that only one wave pulse traversed the contact region at a given time. It is seen from Fig.~\ref{fig:forceTrace} that prior to wave nucleation, $F_T$ builds up due to adherence of the two surfaces. There is a critical tangential force $F_c \simeq 1.6$ N (point $A_1$), at which a single Schallamach wave is nucleated and begins to propagate (point $B$). The corresponding critical interfacial shear stress is $6.5$ kPa. Furthermore, the critical remote strain just prior to the propagation of a wave was estimated to be $\epsilon_c \simeq 0.03$. This is somewhat larger than the shear strain due to slip $|\mathbf{b}|/h_0 \simeq 0.015$. The shear stress relaxes as the nucleated wave traverses the interface and exits the trailing edge of the contact (point $C$). This cycle then repeats with another wave nucleation event. Incomplete readhesion at the interface, caused by the surface wrinkles, results in a reduction of the critical force $(F_c)$ in the cycles that follow the passage of the first wave pulse (point $A_2$). 

Each Schallamach wave pulse produces the same amount of slip (\emph{cf.} Fig.~\ref{fig:surfaceDisp}) irrespective of the sliding velocity. The interface accommodates the imposed $v_s$ by changing the frequency $n$ with which waves are nucleated. The value of $n$ may be obtained from the force trace or the image sequences, both of which are correlated. Fig.~\ref{fig:genFreq} shows the variation of $n$ with $v_s$. The critical velocity $v_c = 150\text{ }\mu$m/s for sliding by Schallamach waves is also marked in the figure. The frequency $n$ is seen to vary linearly with $v_s$ over a large range except very near $v_c$ --- if the best fit line in Fig.~\ref{fig:genFreq} is extended to $n =0$, it intersects the $v_s$ axis at a small negative value. Schallamach waves were observed at velocities very close to, but above, the value $v_c$. The dependence of $n$ on $v_s$ was found to be qualitatively independent of the contact geometry used. 


\section{Analysis and discussion}
\label{sec:analysis}

The high--resolution measurements provide a basis for analyzing nucleation and propagation characteristics of Schallamach waves using simple physical models. Based on the force measurements, we also briefly discuss the stability of homogeneous sliding.

\subsection{Wrinkle pattern during wave nucleation}

The free surface wrinkles, ahead of the lens, result from compression of the elastomer. Correspondingly, an analysis of surface instabilities in a compressed elastic half--space \cite{Biot_ApplSciRes_1963} shows that at a critical compression ratio $\sim 0.5$, the surface is unstable to perturbations of all wavelengths. Hence this cannot explain the observed wavelength pattern. 

However, an analysis of the nucleation stage may be guided by the observed similarity of the surface wrinkle pattern in Fig.~\ref{fig:wrinkles} with that seen in compressed elastic films on soft substrates \cite{CerdaMahadevan_PhysRevLett_2003}. Furthermore, the change in pattern wavelength observed during wave nucleation at large $v_s$, see Fig.~\ref{fig:wrinkles}, appears very similiar to that seen in the longitudinal compression of such a thin film \cite{BrauETAL_NaturePhys_2010}. Motivated by this similarity, we obtain an estimate of the wrinkle amplitude using the model of an elastic thin film on an elastomer substrate, where the elastic properties of the film and the substrate are identical. 

For such a system, the wavelength $\lambda_0$ of the first--appearing wrinkle pattern on the free surface, upon compression, is \cite{Groenewold_PhysicaA_2001}
\begin{equation}
	\lambda_0  = 2\pi \left(\frac{B(1+\nu)(3-4\nu)}{E_s(1-\nu)}\right)^{1/3}
\end{equation}
where $E_s, B, h, \nu$ are Young modulus of the substrate and the film's bending modulus, thickness and Poisson ratio respectively. Using $B = h^3 \, E_s/(12(1-\nu^2))$, $\lambda_0 = 18\text{ }\mu$m from our observations (Fig.~\ref{fig:wrinkles}) and $\nu = 0.45$ for PDMS, we obtain $h \sim 5\text{ }\mu$m, which gives an equivalent \lq film\rq\ thickness. \markthis{A crucial feature is that even though the properties of the thin film and the elastomer substrate are set to be the same, the model used here does not simplify to a regular half--space (Biot's instability). This is because geometric nonlinearity is included in the deformation of the \lq film\rq , while the substrate is assumed to experience small displacement gradients. This approximation is hence consistent, only if the strains are confined to a thin surface layer, as the small value of $h$ \emph{a posteriori} indicates} --- if $h$ were comparable to the elastomer slab height, the surface would prefer to stretch instead of forming wrinkles and buckling, due to energy considerations.

The period doubling with increased compression (Fig.~\ref{fig:wrinkles}(right)) reinforces the thin film analogy \cite{BrauETAL_NaturePhys_2010}. As the compression is increased, the critical compression ratio at which the second (subharmonic) wavelength appears is, to a first approximation, determined entirely by $\nu$ \cite{BrauETAL_NaturePhys_2010}. For $\nu = 0.45$, this occurs at a critical compression ratio $\delta \simeq 0.42$. Correspondingly, the amplitude of the wrinkle pattern is $A \gtrsim h$. With further compression, this amplitude increases with $\delta$. The large amplitude wrinkles that result in residual air pockets in Fig.~\ref{fig:propagation_result} must have an amplitude larger than the thickness $A \gg h$, hence corresponding to large $\delta$. The increased amplitude of the larger wavelength wrinkles on the free surface causes incomplete readhesion after the passage of a single Schallamach wave. This is seen by comparing images of the contact region in the two cases --- low $v_s$ (smaller amplitude, wavelength) in Fig.~\ref{fig:timeFrames} (top, right) and high $v_s$ (larger amplitude, wavelength) in Fig.~\ref{fig:propagationC} --- as well as movie M1 \cite{SuppMat}. It is clear that the number of trapped interfacial air pockets is much greater in the latter case, resulting from an increase in wrinkle amplitude during nucleation. Motion of multiple Schallamach waves hence causes siginificant degradation to the adhesive interface. The value of $\delta$ obtained from this thin film model depends sensitively on $\nu$, for $\nu$ values between 0.45 and 0.49, while the resulting $h$ and $A$ remain roughly the same. 

The model of a thin elastic film on an elastomer substrate thus appears to capture key aspects of the mechanics of nucleation of a single Schallamach wave. \markthis{This analogy hence suggests that by suitable surface treatment (e.g., surface  texturing, pattern impregnation, exposure to ozone) of a very thin (pre-determined) surface layer of thickness $h$, Schallamach waves can be suppressed}. It must be noted, however, that the thin film analogy is based only on wavelength observations and is not fully physically justified. In practice, it is likely that a crease forms on the elastomer surface due to some localized imperfection, as suggested for instance in Ref.~\cite{CaoHutchinson_ProcRoySocA_2011}.

\subsection{Comparison with dislocations and critical stress for propagation}

The observed propagation characteristics of solitary Schallamach waves help expand on the analogy between wave propagation and crystal dislocation glide. Firstly, Schallamach waves are nucleated at a critical stress (point $A_1$, Fig.~\ref{fig:forceTrace}), similar to crystal dislocations. This stress is the compression required for buckling to occur on the elastomer free surface. Secondly, slip at the interface determines an equivalent Burgers vector $\mathbf{b}$ for the Schallamach wave (\emph{cf.} Fig.~\ref{fig:surfaceDisp}) with $|\mathbf{b}|$ independent of $v_s$. This $\mathbf{b}$ can be obtained from surface displacement measurements and shares key characteristics with its dislocation counterpart. Thirdly, when Schallamach waves encounter static dirt particles in the contact region, they are pinned, leaving behind an air pocket separating the two surfaces (see Fig.~\ref{fig:pinning}). This is akin to the well--known pinning of a dislocation line by a solute particle and the resulting residual dislocation loop \cite{Nabarro_CrystalDislocations_1967}. Finally, the nature of the driving force on the wave pulse is similar to the \lq configurational\rq\ Peach--Koehler force on a dislocation --- both the solitary Schallamach wave and an elastic dislocation translate only because their constituent material points move collectively.

\begin{figure}
  \centering
  \includegraphics[scale=0.6]{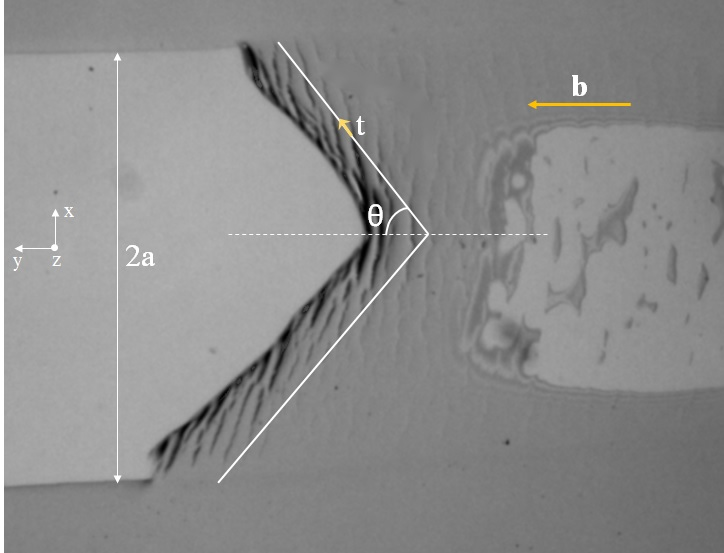}
  \caption{(Color online) Geometry of equivalent line during propagation of Schallamach wave. $2a$ is the width of the contact region.  $\mathbf{t}$ denotes tangent vector along the discontinuity line, which has inclination $\theta$. $\mathbf{b}$ is the Burgers vector. Cylinder lens.}
  \label{fig:dislocationLine}
\end{figure}

For the elastic dislocation model, the (Peach--Koehler) force on the discontiunuity line, arising from applied stress $\sigma$, is given by \cite{Nabarro_CrystalDislocations_1967}
\begin{equation}
\label{eqn:PKforce}
	\tb{F} = \hat{\tb{t}} \times (\tb{\sigma} \cdot \tb{b})
\end{equation}
A typical profile of a single Schallamach wave is reproduced in Fig.~\ref{fig:dislocationLine}. An equivalent dislocation line, subtending an angle $2\theta$ is superimposed over the wave. This line is acted on by a stress state with non-zero components $\sigma_{yz} = \sigma_{zy} = \tau$ and $\sigma_{zz} = \sigma_N$, applied at the contact interface. The resulting surface displacement determines the Burgers vector $\mathbf{b}$ in the $y$ direction. 

If the discontinuity line is displaced by $\delta \tb{r} = \Delta x \, \hat{\tb{x}} \pm \Delta y \, \hat{\tb{y}}$ (for the top and bottom segments), the change in potential energy is given by $\delta V = (\tb{F} \cdot \delta \tb{r})\, L = \frac{4 \tau a|\tb{b}|}{\cos\theta} \, (\sin\theta \Delta x + \sin\theta \Delta y)$, with $L$ the length of the discontinuity line. For the discontinuity to move, the change in the potential energy of the medium must be provided physically by the hysteresis in peeling and readhering of the elastomer surface in the contact region. This energy balance provides an estimate of the critical shear stress needed to propagate a single wave pulse. Assuming $\Delta x \ll a$, the critical force $F_s^c$ required to move a wave pulse is hence given by
\begin{equation}
	F_s^c = \frac{a\,L_c\, \Delta W}{|\mathbf{b}|}
\end{equation}
where $\Delta W$ is the adhesion hysteresis, $2a$ and $L_c$ are the contact width and length respectively. Using $|\mathbf{b}| = 255$ $\mu$m, $2a = 1$ mm, $L_c = 2.5$ cm for the experiments and $\Delta W \simeq 10$ mJ/m$^2$ for PDMS \cite{ChaudhuryETAL_JApplPhys_1996}, $F_s^c$ is estimated to be $0.5$ mN. This is the minimum force needed to propagate a single wave through the contact region. The wave pulse can thus travel through the interface at a much lower stress than that needed for nucleation ($F_c \simeq 1.6$ N). This explains the origin of the drop in tangential force (blue (dark grey) in Fig.~\ref{fig:forceTrace}). As the interface relaxes, the tangential force continues to decrease until it either equals $F_s^c$ or the wave exits the contact region; the latter occurs in Fig.~\ref{fig:forceTrace}. During propagation, a decrease in the normal force $F_N$ (green (light grey) in Fig.~\ref{fig:forceTrace}) results, due to a change in contact size when the wave traverses the contact region. 

\subsection{Slip accumulation in the contact}

Interfacial displacement only results from the passage of single wave pulses; hence the dislocation model can be used to obtain an expression for the strain rate. The interfacial shear strain $\epsilon$ in a time interval $\Delta t$ is due to the passage of $n \Delta t$ parallel dislocation lines. This is given by
\begin{equation}
\label{eqn:strainDislocation}
	\epsilon = \frac{v_s \, \Delta t}{h_0} = \frac{|\mathbf{b}| \, \Delta A}{2 V} \, (n \Delta t)
\end{equation}
where $h_0$ is the height of the elastomer sample, $\Delta A$ the area swept by a single wave in time $\Delta t$, $n$ is the wave generation frequency and $V = 2 a\,h_0 L_c$. The second equality in Eq.~\ref{eqn:strainDislocation} above follows from the expression for shear strain due to glide of a single dislocation \cite{Nabarro_CrystalDislocations_1967}.

As seen in the experiments, $|\mathbf{b}|$ is constant for each wave pulse and also independent of $v_s$. In conjunction with Eq.~\ref{eqn:strainDislocation}, this shows that $n \propto v_s$, consistent with the experimental results of Fig.~\ref{fig:genFreq}. The relation for the strain rate ($\epsilon/\Delta t$) above resembles the Orowan equation for dislocation glide, which relates strain rate in the glide plane to dislocation motion \cite{Nabarro_CrystalDislocations_1967}.

\subsection{Why inhomogeneous sliding?}

Relative motion between two surfaces via propagation of Schallamach waves is part of a larger class of inhomogeneous interface motions constituting stick--slip behavior. We have already mentioned the self--healing slip pulse observed in other sliding systems \cite{BaumbergerETAL_PhysRevLett_2002}. An investigation of forces at the interface provides insight into why such inhomogeneous modes occur.

We use the coordinate system shown in Figs.~\ref{fig:sphereCylinderSchematic} and \ref{fig:dislocationLine}. In this reference, the interface between the lens and elastomer forms part of the $xy$ plane. Before $v_s$ is imposed, the lens and elastomer are in adhesive contact. The corresponding normal force $F_N$ introduces a normal pressure distribution $p(x,y)$ on the elastomer surface. Due to the tangential force $F_T$ (in the $y$-direction), a shear stress $q(x,y)$ also acts on the elastomer surface within the contact region. Typically, the pressure distribution $p(x,y)$ is altered upon the application of $F_T$, but this change is expected to be atmost a few percent \cite{Johnson_ContactMechanics_1987} and can be neglected. 

We define $\mu(x,y) = q(x,y)/|p(x,y)|$ along the interface and let $\mu_0 = F_T/F_N$. The size of the contact region --- diameter of contact circle for spherical lens and width of contact for cylindrical lens --- is $2a$. $K = 4E/3$ with $\frac{1}{E} = \frac{1-\nu_1^2}{E_1} + \frac{1-\nu_2^2}{E_2}$ for the elastic properties $\nu_1, E_1, \nu_2, E_2$ of the elastomer and lens respectively. $R$ is the radius of the lens and $2L$ is the length of the cylinder lens. The exact expression for $\mu(x,y)$ will depend on the contact geometry.

\markthis{When there is no adhesion between the two contacting bodies, the corresponding ratio $\mu(x,y)$ tends to infinity at the edges of the contact. If a Coulomb friction model is assumed, then this implies relative slip locally near the outer edge of the contact zone. Hence a central stick region is postulated inside the contact zone, surrounded by a slip region towards its periphery \cite{Mindlin_JApplMech_1949, Johnson_ContactMechanics_1987}. This cannot be done for an adhesive contact, because the pressure distribution is singular inside the contact, even in the absence of a tangential force. \cite{*[{The tangential loading of adhesive spherical contacts was discussed, using an energy approach, by Savkoor and Briggs}][{}] SavkoorBriggs_ProcRoySocA_1977}}

\markthis{The pressure distribution under static adhesive contact} is axisymmetric when $F_T = 0$, given by \cite{JohnsonETAL_ProcRoySocA_1971, Johnson_ContactMechanics_1987, Barquins_JAdhesion_1988}
\begin{equation}
  p(\xi) = 
  \begin{cases}
    \frac{-1}{2\pi\, a^2(\sqrt{1 - \xi^2})}\left[\frac{F_N}{a} - \frac{K a^2}{R}(3 \xi^2 - 2)\right] & \quad\text{ sphere lens}\\
    \frac{-1}{2\pi\, a^2(\sqrt{1 - \xi^2})}\left[\frac{F_N}{L} - \frac{3\pi K a^2}{8R}(2 \xi^2 - 1)\right] & \quad\text{ cylinder lens}    
  \end{cases}
\end{equation}
where $\xi = r/a$ for sphere lens and $\xi = x/a$ for cylinder lens, and $0 \leq |\xi| \leq 1$. 

When $F_T$ is applied, if the entire contact region moves together, i.e. without relative slip, then the tangential stress on the surface is \cite{Johnson_ContactMechanics_1987}
\begin{equation}
  q(\xi) = 
  \begin{cases}
    \frac{F_T}{2\pi^2a^2} (\sqrt{1-\xi^2})^{-1/2} & \quad\text{ sphere lens}\\
    \frac{F_T}{\pi aL} (\sqrt{1-\xi^2})^{-1/2} & \quad\text{ cylinder lens}\\
   \end{cases}
\end{equation}

In the second expression above for the cylinder lens, the shear stress $q(\xi)$ corresponding to a displacement $u_y$ along the axis of the cylinder ($y$-axis, see Fig.~\ref{fig:sphereCylinderSchematic}), is obtained from the singular integral equation
\begin{equation}
  \frac{\partial u_y}{\partial x} = \frac{-1}{\pi G} \int_{-a}^a \frac{q(s)}{x-s}ds
\end{equation}
for $u_y = \text{constant} = \delta$, i.e. no relative slip between the surfaces.

The ratio $\mu(\xi) = q(\xi)/p(\xi)$ is given by
\begin{equation}
\label{eqn:mu_values}
  \mu(\xi)/ \mu_0 = 
  \begin{cases}
      |\pi(1 - \frac{Ka^3}{F_N R}\,(3\xi^2-2))|^{-1} & \quad\text{ sphere lens}\\
      |8 - 3\pi \frac{KLa^2}{F_N R}\,(2\xi^2-1)|^{-1} & \quad\text{ cylinder lens}\\
  \end{cases}
\end{equation}

This expression for $\mu(\xi)/\mu_0$ has a singularity at $\xi_c = \left(\frac{2}{3} + \frac{F_N R}{3 K a^3}\right)^{1/2}$ and $\xi_c = \left(\frac{1}{2} + \frac{4 F_N R}{3 \pi K L a^2}\right)^{1/2}$ for the spherical and cylinder lens geometries, respectively. This $\xi_c$ corresponds to the point at which the normal stress $p(\xi)$ changes sign from compressive to tensile inside the contact region. It is interesting to note that $\xi_c$ is independent of the applied tangential force $F_T$ and always lies within the contact region $0 \leq \xi \leq 1$. Even if the singularity inherent in Eq.~\ref{eqn:mu_values} is replaced by a large finite value, the fact that no relative homogeneous slip occurs at the interface ensures that, locally, the static friction coefficient $\mu > \mu(\xi)$.

The implication of the above expression for $\mu(\xi)$ can be stated as follows, following Ref.~\cite{RanjithRice_JMechPhysSolids_2001}. For frictionless contact between \lq sufficiently\rq\ dissimilar materials, as with the lens--elastomer system in our experiments, the generalized Rayleigh wave \cite{Weertman_JMechPhysSolids_1963} does not exist. In this case, there is a finite value of $\mu_0$ ($\lesssim 1$) above which homogeneous sliding is unstable to perturbations of all wavelengths.  

Assuming a Coulomb model in the experiments and using the force values in Fig.~\ref{fig:forceTrace}, the effective static friction coefficient, $\mu_0 = F_T/F_N$, is very high. Prior to nucleation, the surfaces are stationary even when $F_T  \gg F_N$, until at $F_T = F_c$, a Schallamach wave is nucleated. This relaxes the stresses, thereby lowering the value of $\mu_0$. One can thus infer that, for the lens--elastomer system, homogeneous interfacial sliding is unstable for large values of $\mu_0$. Since $\mu(\xi)$, for $|\xi|<1$ is always larger than a finite value, locally homogeneous sliding is unstable. Furthermore, the singularity point $\xi_c$ in Eq.~\ref{eqn:mu_values} occurs inside the contact region (and not at the edge, as in the purely elastic case), interfacial stick and slip regions cannot exist. This also rules out partial homogeneous interfacial motion. Thus, inhomogeneous sliding modes are very likely to occur in cases involving adhesion. 


\section{Conclusions}
\label{sec:conclusions}

The nucleation and propagation of isolated Schallamach waves in an adhesive elastomer contact has been studied \emph{in situ} using high--speed imaging. These two phases of inhomogeneous sliding were observed and characterized using spherical and cylindrical lens contacts. The former enabled observation of nucleation features such as the formation of wrinkles and their subsequent evolution. The latter was conducive for isolating and analyzing the dynamics of solitary Schallamach waves. Based on characterization of the individual wave properties, a Burgers vector, analogous to dislocations, was established. Pinning of Schallamach waves by static dirt particles and existence of critical nucleation force were also demonstrated, all of which have analogues in dislocation glide. Simple analytical considerations of the contact stresses also provide clues as to why inhomogeneous sliding modes via Schallamach waves may be preferred in adhesive contact systems.

\section*{Acknowledgements}
This work was supported in part by US Army Research Office grant W911NF-12-1-0012 and NSF grants CMMI 1234961 and 1363524. Insightful discussions with Dr. J. Hanna and Dr. D. P. Holmes (Virgina Tech.); and Dr. N. K. Sundaram (Indian Institute of Science) are gratefully acknowledged. We thank the referees for constructive feedback on the manuscript.


\appendix

\section{Image processing methods}
\label{sec:appendix}

The high--speed image sequence obtained from the experiments were analyzed in order to obtain pixel--level velocities. \markthis{Even though the elastomer appears transparent, very small opaque features (such as embedded minute dust particles) always exist, providing contrast for tracking purposes}. The estimated inter--frame pixel displacements are proportional to the local pixel velocity $\tb{v}_p(x,y)$ because the time between frames is constant.

In order to do this, the local image intensity $I(\mathbf{x})$ at each image point $\mathbf{x} \equiv (x,y)$ is approximated locally by a quadratic polynomial, i.e. $I(\mathbf{x}) = \mathbf{x}^T \mathbf{A} \mathbf{x} + \mathbf{b}^T \mathbf{x} + c$, with coefficients $\mathbf{A}, \mathbf{b}, c$ determined by a weighted least squares fit to intensity values in the neighborhood of $\mathbf{x}$. These coefficients are computed for each pixel in the image. If the image intensity is convected with the velocity field, $I(\mathbf{x} + \mathbf{d}, t+ dt) = I(\mathbf{x}, t)$ for a pixel $\mathbf{x}$ at time $t$, translated by $\mathbf{d}$ to the next frame $dt$ seconds later; $\mathbf{d}(\mathbf{x})$ is obtained by comparing these intensities.

From a practical point of view, this results in significant noise in the displacement field. This is overcome by assuming that $\mathbf{d}$ is slowly varying and performing a weighted average over pixels in a window. Furthermore, to minimize estimation error for large displacements, \emph{a priori} estimates are used for each frame by performing the displacement estimation at multiple length scales \cite{Farneback_ImageAnalysis_2003}. The displacement fields are first calculated at a coarse scale, for large blocks of the image. Subsequently, the chunk size is reduced and the coefficients above are iteratively calculated with information from the previous scale as an \emph{a priori} estimate. This method hence works even for large displacements. The entire scheme was implemented by combining custom code with functions from the OpenCV code library \cite{opencv_library}.

For the analyses reported here, a window size of 15 pixels was found to give the best tradeoff between noisy data and a blurred velocity field. Three successive scales were used for the \emph{a priori} estimate and at each scale the image size was halved. To refine the displacement estimate, the algorithm was iterated 3 times at each scale. For estimating $\mathbf{A}, \mathbf{b}, c$, a neighborhood size of 5 pixels was found to give good results for the least squares fit. 


\bibliography{manuscript}
\end{document}